\documentclass[pre,twocolumn,twoside,byrevtex,superscriptaddress,floatfix,secnumarabic,amssymb,nobibnotes,aps,pr,titlepage]{revtex4-1}

\usepackage{settings}
\usepackage[utf8]{inputenc}
\usepackage[T1]{fontenc}
\usepackage{amsmath}
\usepackage{ragged2e}
\usepackage[hidelinks]{hyperref}
\hypersetup{breaklinks=true}
\DeclareMathOperator*{\argmax}{arg\,max}
\DeclareMathOperator*{\argmin}{arg\,min}

\makeatletter
\renewcommand\@make@capt@title[2]{%
  \@ifx@empty\float@link{\@firstofone}{\expandafter\href\expandafter{\float@link}}%
   {\textbf{#1}}\@caption@fignum@sep#2\quad
}%
\makeatother
\usepackage{graphicx}
\usepackage{booktabs}
\usepackage{rotating}
\usepackage{adjustbox}
\usepackage{url}
\usepackage{hyperref}
\usepackage[table]{xcolor}
\rowcolors{2}{gray!25}{white}
\usepackage{lineno}
\newcommand{\UVM}{ Computational Story Lab, Vermont Complex Systems Center,
 MassMutual Center of Excellence for Complex Systems \& Data Science, University of Vermont}
\newcommand{\mathDept}{Department of Mathematics \& Statistics, University of Vermont}

\newcommand{\umass} {Department of Biostatistics \& Epidemiology,
 School of Public Health \& Health Sciences,
 University of Massachusetts at Amherst}
 
\newcommand{\MM}{MassMutual Data Science}

\begin{document}

\title{The sleep loss insult of Spring Daylight Savings in the US \\ 
is absorbed by Twitter users within 48 hours}

%% By Monday morning, activity is back tonormal, suggesting roughly an hour of sleep lost

%% \title{Spring Daylight Savings as a natural experiment in sleep lost to Twitter}

%\title{Tweeting instead of sleeping: Spring Daylight Savings as a natural experiment in sleep l

%\title{\normalfont Stop tweeting and start sleeping: spring Daylight Savings as a natural experiment in sleep loss}

%\title{\normalfont Sleep! Don't tweet.: spring Daylight Savings as a natural experiment in sleep loss}

%\title{\normalfont How much sleep do we lose to spring Daylight Savings? A population-scale analysis of tweets suggests 45 minutes.}

%\title{\normalfont Estimating sleep opportunity lost during ``Spring Forward'' daylight savings events from tweets}

\author{Kelsey Linnell}
\email{klinnell@uvm.edu}
\affiliation{\UVM}
\affiliation{\mathDept}

\author{Thayer Alshaabi}
\affiliation{\UVM}

\author{Thomas McAndrew}
\affiliation{\umass}

\author{Jeanie Lim}
\affiliation{\MM}

\author{Peter Sheridan Dodds}
\affiliation{\UVM}
\affiliation{\mathDept}

\author{Christopher M. Danforth}
\email{chris.danforth@uvm.edu}
\affiliation{\UVM}
\affiliation{\mathDept}

%\author[UVM]{Thayer Alshaabi}
%\author[Umass]{Thomas McAndrew}
%\author[MM]{Jeanie Lim}
%\author[UVM,math]{Peter Sheridan Dodds}
%\author[UVM,math]{Christopher M. Danforth}

% \address[UVM]{Computational Story Lab, Vermont Complex Systems Center,
% MassMutual Center of Excellence for Complex Systems \& Data Science, University of Vermont}

% \address[math]{Department of Mathematics \& Statistics, University of Vermont}

% \address[Umass]{Department of Biostatistics \& Epidemiology,
% School of Public Health \& Health Sciences,
% University of Massachusetts at Amherst}
% ]
% \address[MM]{MassMutual Data Science}
%\end{frontmatter}

\begin{abstract}
\noindent \textbf{Abstract:} 
Sleep loss has been linked to heart disease, diabetes, cancer, and an increase in accidents, all of which are among the leading causes of death in the United States. Population-scale sleep studies have the potential to advance public health by helping to identify at-risk populations, changes in collective sleep patterns, and to inform policy change. 
Prior research suggests other kinds of health indicators such as depression and obesity can be estimated using social media activity.
However, the inability to effectively measure collective sleep with publicly available data has limited large-scale academic studies.
Here, we investigate the passive estimation of sleep loss through a proxy analysis of Twitter activity profiles.
We use ``Spring Forward'' events, which occur at the beginning of Daylight Savings Time in the United States, as a natural experimental condition to estimate spatial differences in sleep loss across the United States. 
On average, peak Twitter activity occurs roughly 45 minutes later on the Sunday following Spring Forward. 
By Monday morning however, activity curves are realigned with the week before, suggesting that at least on Twitter, the lost hour of early Sunday morning has been quickly absorbed.

\end{abstract}

\maketitle

\section{Introduction}

The American Academy of Sleep Medicine recommends adults sleep 7 or more hours per night~\cite{Watson2015al}. 
However, studies show only 2/3 of adults sleep for this length of time consistently. In 2014, the Centers for Disease Control and Prevention's (CDC's) Behavioral Risk Factor Surveillance System suggested that between 28\% and 44\% of the adult population of each state received less than the recommended 7 hours of sleep~\cite{CDC}.
Despite the scientific consensus that adequate sleep is essential to health, many adults are sleeping less than $7$ hours a night on average---a state referred to as \textit{short sleep}.
Results from the most recent National Health Interview Survey determined that since 1985, the age-adjusted average sleep duration has decreased, and the percentage of adults who experience short sleep, on average, rose by 31\%~\cite{ford}.

%P2-the implications of this problem on the public
Because adequate sleep is necessary for optimal cognition, short sleep is adverse to productivity and learning, and reduces the human capacity to make effort- related choices such as whether to take precautionary safety measures ~\cite{Althoff,curcio,engle}.
Short sleep's impact on human cognition is harmful in the workplace, and poses a pronounced and distinct threat to public safety when operating a vehicle~\cite{rosekind,dean,Owens,Andersen}.
Short sleep is linked to increased risk of serious health conditions, including heart disease, obesity, diabetes, arthritis, depression, strokes, hypertension, and cancer~\cite{goran,CDC,walker,nagai},
and a recent study found that disrupted sleep %\tcm{is this the same as short sleep? if not delete}
is also associated with DNA damage~\cite{Cheung}. 
The link between sleep loss and cancer is so strong that the World Health Organization has classified night shift work as ``probably carcinogenic to humans''~\cite{Reuters}.
Socio-economic status is positively correlated with quality of sleep~\cite{patel, chattu, ruiter, curtis}.
Due to such detrimental effects, and high prevalence among the population, insufficient sleep accounts for between \$280 and over \$400 billion lost in the United States every year~\cite{hafner}.

% P4-It is tricky to measure short sleep on a large scale, and at the same time,accurately.
Accurately measuring short sleep in a large population is difficult, and there is often a trade-off between accuracy and the size of the study.
Polysomnography---considered the most accurate way to measure sleep---can only measure an individual's sleep patterns in a controlled laboratory setting~\cite{bianchi,douglas}.
Large studies have relied on participants recording their own sleep, but suffer from reporting bias~\cite{Harvey,selfreport,CDC}. %\kml{could cite kobyashi and gradisar as examples of large scale studies?}

%P5-
Wearable technology can measure short sleep at the population scale, and has the potential to measure short sleep accurately enough to study its association with adverse health risks~\cite{Althoff,bianchi, marino}.
One recent large sleep study enrolled 31,000 participants and used sleep data from wearable devices along with participant's interactions with a web based search engine to compare sleep loss and performance~\cite{Althoff}.
The authors ~\cite{Althoff} showed that measurements of cognitive performance (including keystroke and click latency) vary over time, follow a circadian rhythm, and are related to the duration of participant's sleep, results that closely mirrored those from laboratory settings and validated their methodology. 

While promising in the long run, present 
studies that use wearable devices have limitations.
To infer from wearables that individuals are sleeping, data must first go through a pipeline of preprocessing, feature extraction and classificiation.
The pipeline for processing sleep data is typically proprietary and dependent on the specific wearable used, and changes to how data is processed can impact results~\cite{Roomkham}. Moreover, validation studies have yet to explore the effectiveness of these devices across genders, ages, culture, and health \cite{Roomkham}. 

Social media may be an alternative way to measure sleep disturbances in a large population, for example by studying the link between screen time and sleep~\cite{Leypunskiy,Christensen}.
Past work has found a correlation between sustained low activity on Twitter and sleep time as measured by conventional surveys, and these results were validated against data collected from the CDC on sleep deprivation~\cite{Leypunskiy}.
Other work has shown evidence of an increase in a user's smart phone screen time as being associated with an increase in short sleep~\cite{Christensen}. 
Other mental and physical characteristics have been measured from sociotechnical systems. 
Several instruments developed by members of our research group
including the Hedonometer~\cite{Hedo}, which measures population sentiment through tweets, and the Lexicocalorimeter~\cite{Lexicocalorimeter}, which measures caloric balance at the state level, have demonstrated an ability to infer population-scale health metrics from Twitter data. 
Twitter data has also been used to identify users who experience sleep deprivation and study the ways their social media interactions differ from others~\cite{mciver}.

In urban, industrialized societies where social timing is synced to clock time, Daylight Savings- a biannual sudden upset to clock time- creates behavioral stability across seasons~\cite{martin_dst,martin_scand}. Past work has used Daylight Savings as a natural experiment to show that a one hour collective sleep loss event has large and quantifiable effects on health, safety, and the economy~\cite{sandhu,sipila,varughese,martin_traffic},
with two striking findings being a
one day increase in heart attacks by 24\% and a loss of \$31 billion on the NYSE, AMEX, and NASDAQ exchanges in the United States~\cite{sandhu,kamstra}.

We hypothesize here that sleep loss is measurable in behavioral patterns on Twitter, and changes in population-scale sleep patterns due to Spring Forward can be observed through changes in these behavioral patterns. 
In what follows, we first outline our methodology for estimating sleep loss from tweets, describing the data and study design. We then visualize and describe the results before concluding with a discussion of limitation and implications.

\begin{figure*}[ht!]
    \centering
    \includegraphics[width=\textwidth]{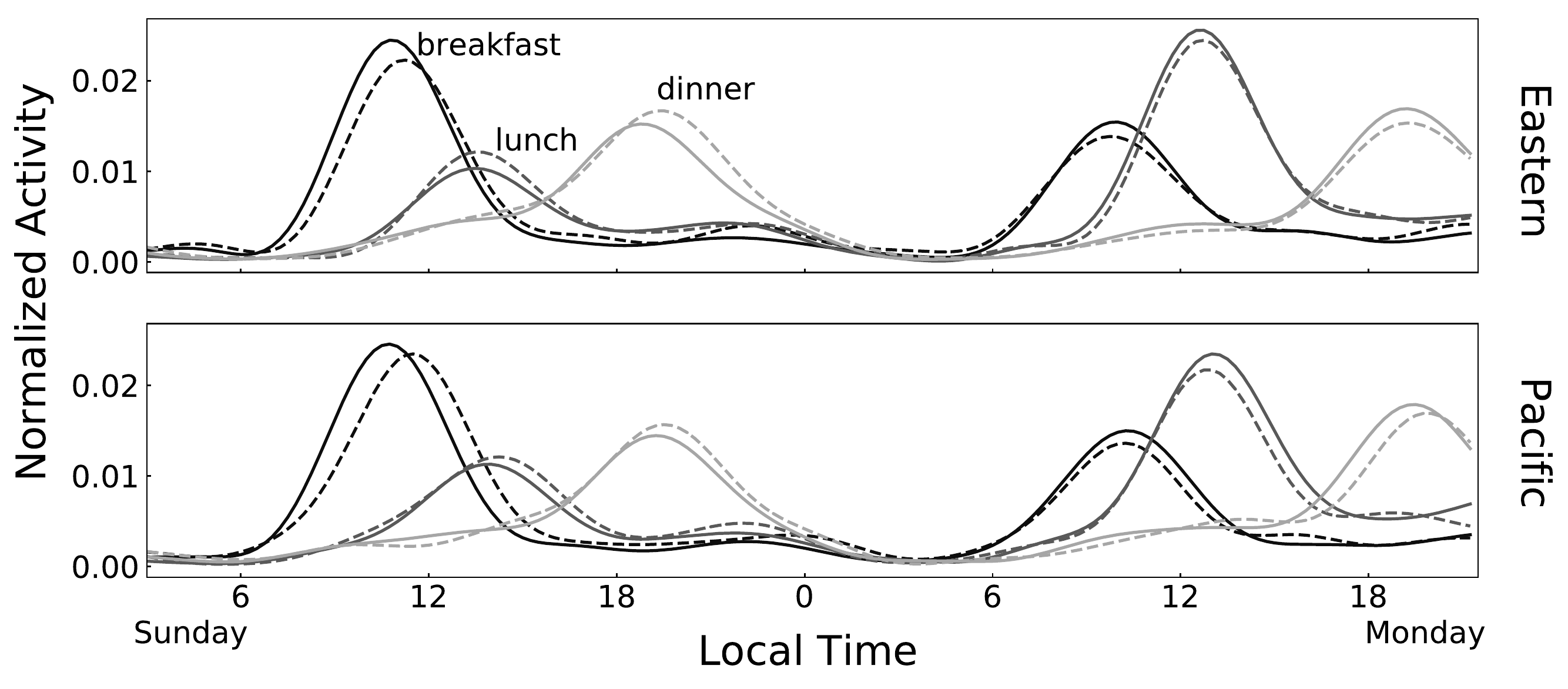}
    \caption{ \textbf{Diurnal collective attention to meals quantified, by normalized usage of the words `breakfast', `lunch', and `dinner' for states observing Eastern Time (top) and Pacific Time (bottom), for the weeks before (solid), and of (dashed) Spring Forward.}
    The x-axis represents the interval between 3 a.m. Sunday and 9 p.m. Monday local time. 
    Counts for tweets containing each individual word were tallied in 15 minute increments, normalized by the total number of tweets mentioning that word, and smoothed using Gaussian Process Regression. 
    Each day has a clear pattern for frequency of meal name appearance in tweets, with the peak for breakfast, lunch, and dinner occurring in the respective order of the meals themselves.
    For each of the meals, we observe a slight forward shift in the peak following Spring Forward, suggesting that meals are taking place later than usual on the corresponding Sunday.
    By Monday, the peak for each meal name appears to be aligned with the week before, with the exception of 'dinner' on the west coast, which is still a bit later.} 
    \label{fig:mealfig}
\end{figure*}

\section{Methods}

\subsection*{Data}

We collected a 10\% random sample of all public tweets---offered by Twitter's Decahose API---for Sundays and Mondays in the four weeks leading up to, the week of, and the four weeks following Spring Forward events during the years 2011-2014.
Spring Forward is defined as the instantaneous clock adjustment from 2 a.m. to 3 a.m. on the second Sunday of March each year.
We included tweets in the study if the user who created the tweet reported living in the U.S.\ in their bio, or if the tweet was geo-tagged to a GPS coordinate within the U.S.~\cite{gray}. 
With these conditions, we ended up selecting approximately 7\% of the messages in the Decahose random sample for analysis~\cite{Decahose}.

Twitter provided the time-zone from which each message was posted during the period from 2011 to 2014~(for privacy purposes, Twitter discontinued publication of time zone information in 2015).
We used the time-zone to determine the local time of posting for each tweet.
We binned tweets by 15 minute increments according to the local time of day they were posted.

\subsection*{Experimental setup}

To estimate behavioral change associated with Daylight Savings, we partitioned tweets into various groups, primarily a ``Before Spring Forward'' (BSF) group and a ``Spring Forward'' (SF) group.
To establish a convenient `control' pattern of behavior, all tweets posted on any of the four Sundays before the Spring Forward event were classified as ``Before Spring Forward'' tweets. We classified the `experimental' set of tweets posted on the Sunday coincident with the Spring Forward event as ``Spring Forward''.
The above classification created, for every year, a 4:1 matching of before to week of Spring Forward activity. 
We analyzed tweets posted 1-4 weeks following Spring Forward separately to quantify relaxation to the original behavior.

\subsection*{Analysis}

We binned tweets by time in 15 minute intervals starting at the top of the hour, and normalized their frequencies by dividing by the total number of tweets posted on the corresponding day. 
In this way, we establish a discrete description of the posting volume over the course of a typical 24-hour period.

We averaged the Before Spring Forward tweets over the four Sundays, and the four years as follows:
\begin{equation*}
   T_{\text{BSF}}(k)= \left(4 \times 4\right)^{-1} \sum_{Y=2011}^{2014}~\sum_{\text{S}=1}^{4}\frac{C_{YS}(k)}{C_{YS}}, 
\end{equation*}
where $C_{YS}(k)$ is the number of tweets in the $k^{th}$ 15 minute interval of the $S^{th}$ Sunday of year $Y$, $C_{YS}$ is the total number of tweets posted on that Sunday and year, and $T_{\text{BSF}}(k)$ is the average fraction of tweets posted in the $k^{th}$ 15 minute interval of a Sunday prior to Spring Forward, 

We also noramlized the Spring Forward tweets 
against daily activity:
\begin{equation*}
   T_{\text{SF}}(k)= \left(4\right)^{-1} \sum_{Y=2011}^{2014}\frac{C_{Y}(k)}{C_{Y}}. 
\end{equation*}

To reduce noise that could depend on our choice of bin size and spatial scale, we smoothed normalized tweet activity using Gaussian Process Regression (GPR)~\cite{rasmussen2003gaussian, scikitlearn}.
We fit a GPR with a squared exponential kernel and characteristic length scale of 150 minutes (a total of 10 bins of size 15-minutes) to normalized tweets.
We chose a characteristic length of 150 minutes 
for consistency with previous work~\cite{Leypunskiy}. 
Tikhonov regularization with an $\alpha$ penalty of 0.1 was included when finding weights $\omega_{k}$ to prevent overfitting~\cite{scikitlearn}.
GPR yielded a smooth \textbf{behavioral curve}, $B(t)$, of the functional form:
\begin{equation*}
    B(t)= \sum_{k=1}^{96} \omega_{k} \exp \left[-\frac{1}{2} k\left(\frac{t}{150},\frac{t_k}{150}\right)^2\right],
\end{equation*}
where $\omega_k$ is a weight determined by the regression process, $k$ is the squared-exponential kernel~(commonly called a radial basis), $t$ is the time in minutes since midnight (00:00), and $t_k$ is the $k^{th}$ 15 minute interval of the day, i.e. $t_5$ corresponds to 75 minutes past midnight, or 1:15 a.m. The sum to 96 refers to the number of 15 minute intervals in a single 24 hour period.

\begin{figure*}[ht!]
    \centering
    \includegraphics[width=\textwidth]{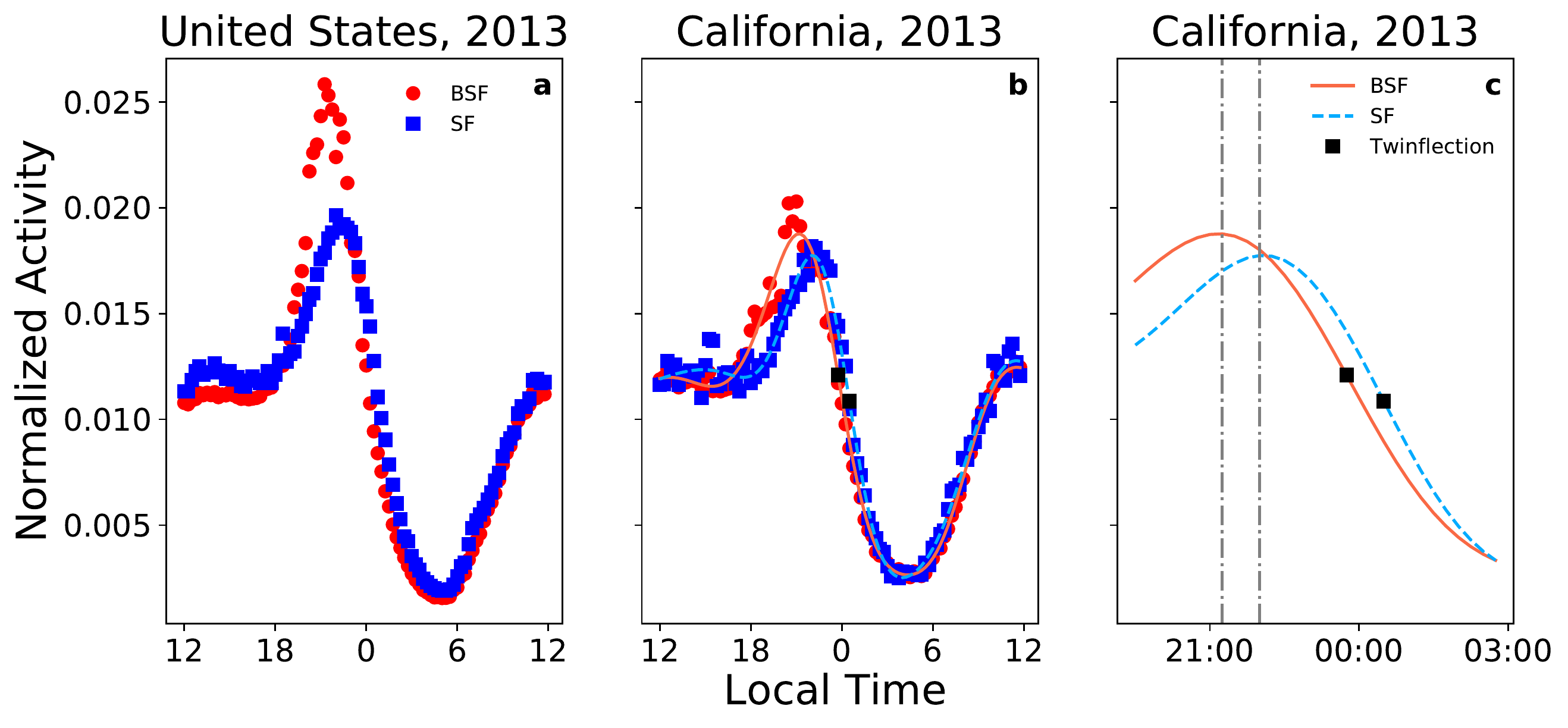}
    \caption{
    \textbf{Twitter activity behavioral curves $B(t)$.} \textbf{(a)} Normalized count of tweets posted from a location within the United States between 12 p.m. Sunday and 12 p.m. Monday before (red) and the week of (blue) the 2013 Spring Forward Event. The time recorded for the tweet is that local to the author. Though the pattern of behavior is preserved following Daylight Savings, peak activity is translated forward in time. \textbf{(b)} The same plot, with location of tweet origin restricted to the state of California. California is the state for which we have the most data, and therefore the most representative behavior profile after smoothing with Gaussian Process Regression (lines). We note that figure \ref{fig:statecurves} shows behavioral curves for all states. \textbf{(c)} The smoothed behavioral pattern for California during the hours of 9 p.m. to 3 a.m. Pacific Time. Activity peaks are denoted by vertical dashed lines, and twinflection points are marked by squares. To estimate the behavioral shift in time, we compute the distance along the temporal axis between these pairs of lines/points. California's BSF peak is 30 minutes earlier than the SF peak. %\tcm{Why California? Because of Katy Perry's song? Because this showed the largest shift in peak? it would be neat to see two examples, one where the late shift is very pronounced and a counter example (if one exists) where there is really no shift in peak activity-KML: see paragraph, just an example state. CD: Since Figure 5 shows every state, I think it is ok to just show a single one here. Made note of this in the caption so the curious can skip ahead.}
    }
    \label{fig:behaviorcurves}
\end{figure*}

We generated behavioral curves $B(t)$ for the BSF and SF groups by state, and for the U.S. in aggregate. 
To estimate behavioral change induced by a Spring Forward event, we calculate two quantities from the behavioral curves: (i) the time of peak activity and (ii) the time of the inflection point between the peak and trough.
The inflection point is referred to as a `twinflection' point, and represents a point of diminishing losses in Twitter activity for the night.
Peak shift is defined as:
\begin{equation*}
    \argmax_t{ \left\{B_{\text{SF}}(t) \right\}}-\argmax_t{ \left\{B_{\text{BSF}}(t)\right\}}
\end{equation*}
and twinflection shift is defined as:
\begin{equation*}
    \argmin_{t\in N}{ \left\{B_{\text{SF}}'(t) \right\}}-\argmin_{t\in N}{\left\{B_{\text{BSF}}'(t)\right\}},
\end{equation*}
where $N=\{ t : \argmax_t{B(t)}<t<\argmin_t{B(t)} \}$.
We were able to reliably measure peak activity and twinflection because behavioral curves exhibited a consistent diurnal wave structure: a rise in the evening corresponding to peak Twitter posting activity, followed by a trough during typical sleeping hours, and a plateau throughout the day. 

We measured the loss of sleep opportunity by calculating the peak and twinflection times for the four weeks Before Spring Forward and the week of Spring Forward itself. We then characterize differences between the BSF and SF measures for each state, and for the total U.S., as a proxy for sleep loss.  %\tcm{Maybe one more sentence that lays out the comparison you'll make and why. CD: Good idea, done.}

\section{Results}
\noindent

Our overall finding is that peak Twitter activity occurs roughly 45 minutes later on the Sunday evening immediately following Spring Forward, with this shift varying among states. 
By Monday morning, activity is back to normal, suggesting that the hour of sleep lost is overcome, at least on Twitter, within 48 hours.

In Fig~\ref{fig:mealfig}, we plot $B(t)$ for the subset of posts containing the words `breakfast', `lunch', and `dinner' for the period beginning 3 a.m. on Sunday and ending 9 p.m.
on Monday, both before (solid) and after (dashed) Spring Forward events. 
These curves were constructed for states observing Eastern Time (top row) and Pacific Time (bottom row).

\begin{figure*}[ht!]
    \centering
    \includegraphics[width=\textwidth]{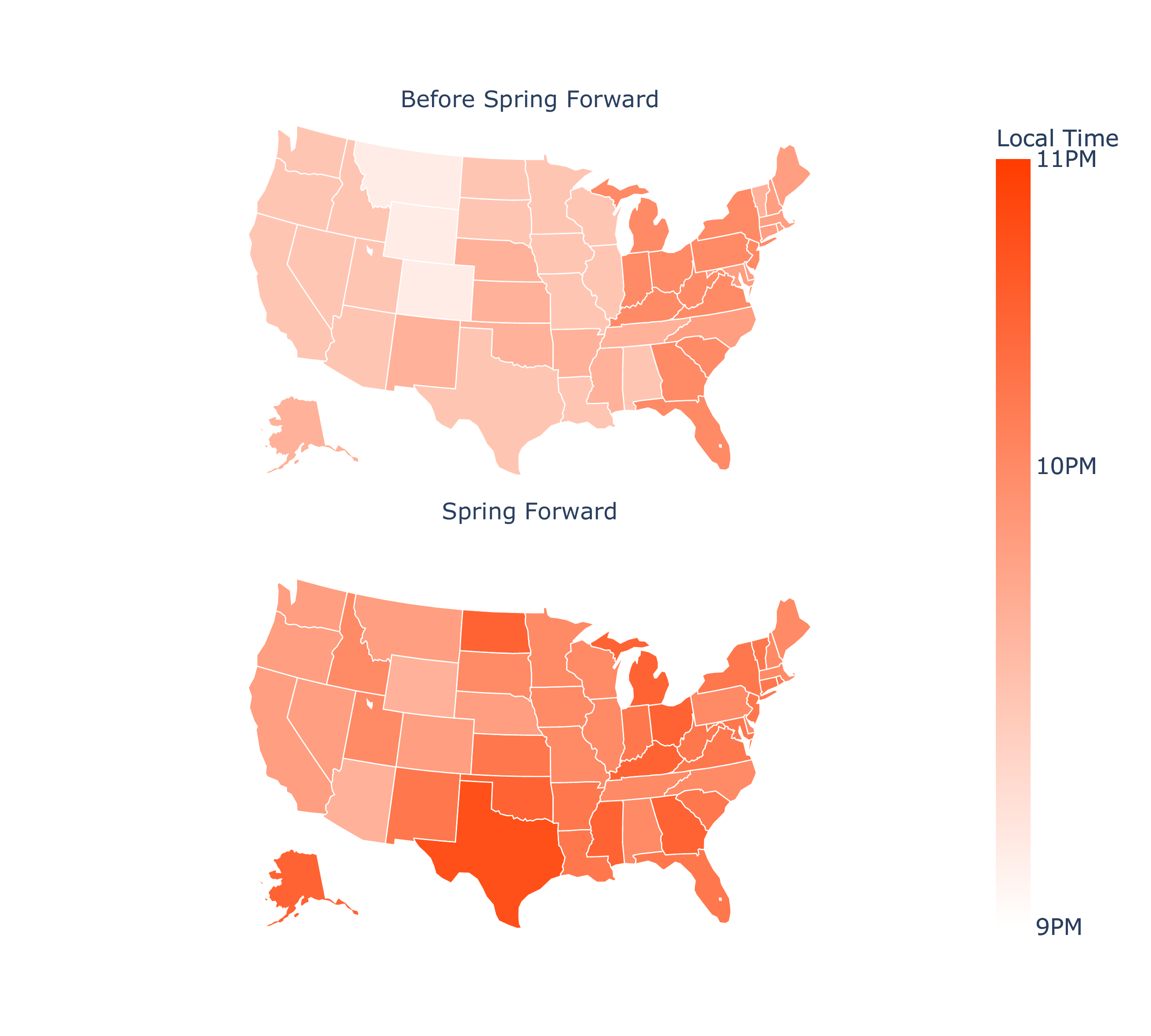}
    \caption{\textbf{
    Time of peak Twitter activity on Sunday night for each state before (top) and after (bottom) Spring Forward for the four events observed between 2011 and 2014.}
    Before Spring Forward, the time of peak activity occurs around 10 p.m. in the Eastern Time Zone, and around 9:30 p.m. for the rest of the country. 
    After Spring Forward, peak Twitter activity occurs between 0 and 90 minutes later for each state. 
    Texas has the latest peak at 11 p.m. local time, a shift of 90 minutes forward compared with prior Sundays.
    Pennsylvania, Hawaii, and Washington D.C. are the only states with no observed change in peak time. We note again that the BSF estimates are based on the aggregation of four Sundays prior to Spring Forward, while the ASF estimates are based on the Sunday coincident with Spring Forward, and are therefore estimated using roughly 1/4 the data.}
    \label{fig:peakmap}
\end{figure*}

Meal-related language reveals a daily pattern of behavior in which peak volume occurs around the time that meal typically takes place.
On an average Sunday, breakfast is most mentioned at 11 a.m., lunch at 1:45 p.m., and dinner at 7 p.m. in Eastern Time Zone states (see Fig~\ref{fig:mealfig}). On the average Monday, breakfast mentions peak at 10:15 a.m., lunch peaks at 1 p.m., and dinner at 8 p.m.  
Breakfast is mentioned nearly twice as often on Sunday than on Monday. 
Lunch shows the opposite trend, doubling on Monday in comparison to Sunday.
There is essentially no discussion of meals during the period from 2 a.m.-4 a.m.
These plots also exhibit a small forward shift in time following Spring Forward, suggesting that each meal was tweeted about, and probably eaten, later in the day on Sunday. 
The effect disappears by Monday.
%\tcm{You mention above a difference between the frequency of Sunday and Monday mentions of breakfast lunch and dinner. I didn't see a follow-up sentence describing any differences in the ratio of the frequency of BLD before and after SF? % CD: the sentence a few lines up is good. The ratio in frequencies is very close to 1 looking at the figure, don't think it is worth mentioning}

Broadening from messages mentioning specific meals to all messages, daily activity plots of $B_{BSF}$ and $B_{SF}$ reveal a regular diurnal pattern of behavior that is consistently shifted forward in time the evening following Spring Forward events. 
Figure ~\ref{fig:behaviorcurves} shows this shift for the year 2013, but the results were similar for other years.
Panel (a) suggests overall activity across the U.S. peaks around 10 p.m. on Sundays before Spring Forward (red circles), and experiences a minimum around 5am. 
The peak shifts approximately 45 minutes later on the Sunday of Spring Forward (blue squares) before synchronizing again by early morning Monday. 
In panel (b) California is used as an illustrative example of these patterns existing at the state level, and the smooth behavioral pattern constructed using Gaussian Process Regression. 
The pattern is similar to that observed for the entire country, with the exception of a slightly reduced amplitude.
Twinflection points are illustrated by black squares in panels (b) and (c).
%\tcm{same question about CA as in Fig2 caption. %CD I think ok to just point to figure 5}

Figure \ref{fig:behaviorcurves} demonstrates evidence that there is a shift in the peak time spent interacting with Twitter on Sunday evening following Spring Forward, relative to prior Sundays.
Given the absence of a corresponding delay in interaction Monday morning, we infer an increase in sleep loss experienced on Sunday night.

\begin{figure*}[ht!]
   \centering
     \includegraphics[width=\textwidth]{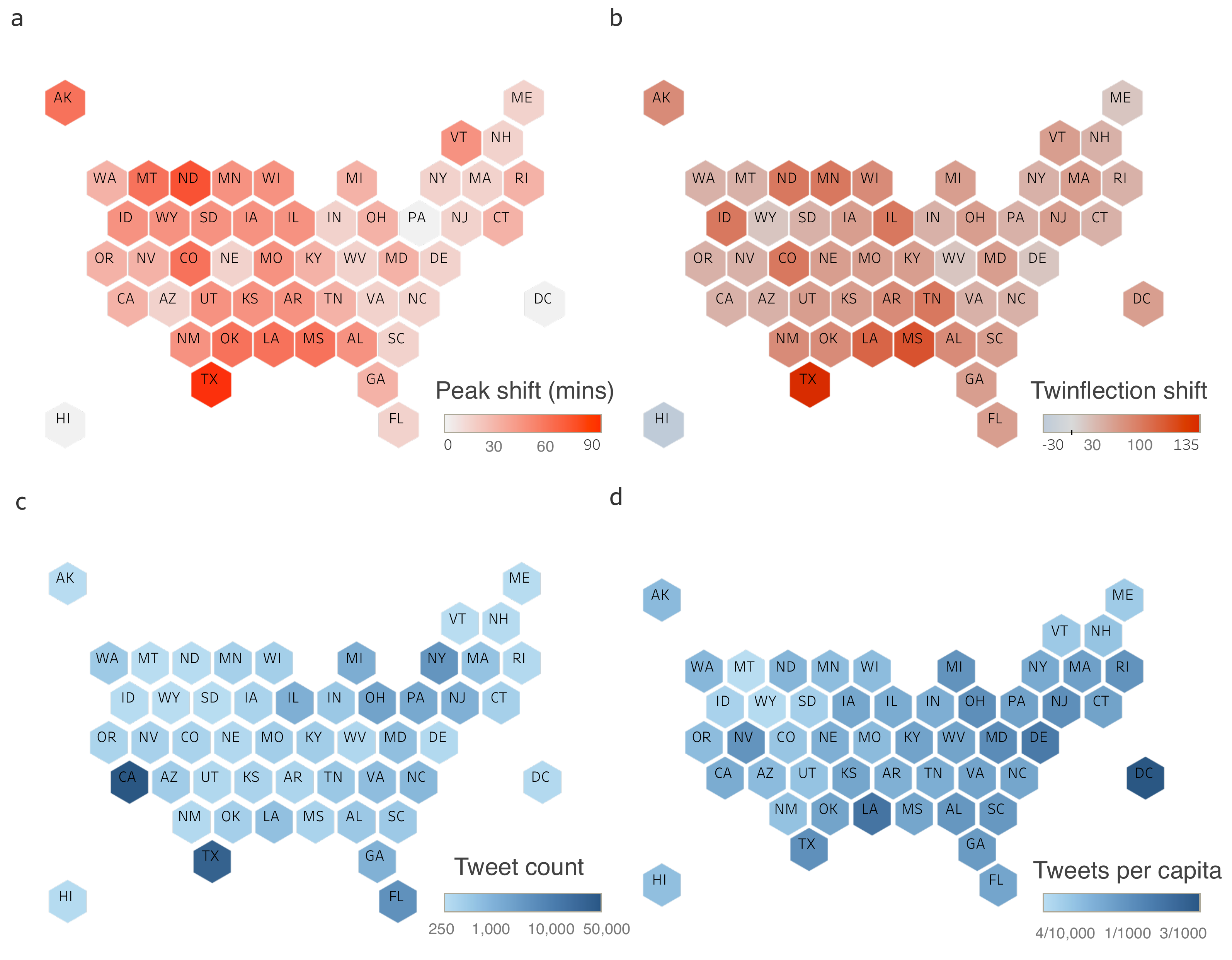}
     \caption{\textbf{The magnitude of Twitter behavioral shift following a Spring Forward event, averaged for the four years from 2011 to 2014.} (a) Shift measured using behavioral curve peaks, the difference between the pair of maps in Figure \ref{fig:peakmap} (bottom minus top). Texas is estimated to have experienced the greatest time shift. The effect of Spring Forward is more pronounced in the South, and center of the country. No effect is measured for Hawaii. 
     \textbf{(b)} The same map, but with measurements calculated using twinflection shift instead. The states most affected are Texas and Mississippi, where the shift was 135 and 105 minutes respectively. Hawaii is the only state estimated to have a negative shift (30 minutes). Twinflection shift produces similar spatial results to peak shift, with more exaggerated shift estimates.
     \textbf{(c)} The number of tweets posted from each state in the period after Spring Forward. California and Texas both contributed over 40,000 tweets, while Alaska, Hawaii, Idaho, Wyoming, Montana, North Dakota, South Dakota, and Vermont each produced less than 1,000 tweets.
     \textbf{(d)} The density of data used to establish the experimental pattern of behavior, as measured by tweets per capita. This measurement reflects the ability of the data to capture the behavior of the tweeting population of each state. While Idaho, Wyoming, Montana and South Dakota have relatively little data compared to their populations, the remaining states have similar data density, with somewhere between one and three tweets per thousand residents. Note: both panels (c) and (d) use logarithmically spaced colorbars.
    \label{fig:fourmaps}}
\end{figure*}

To explore the spatial distribution of the behavioral changes induced by Spring Forward, in Fig.~\ref{fig:peakmap} we map the time of peak Twitter activity on Sunday night for each state before (top) and the week of (bottom) Spring Forward, averaged across the years 2011-2014. 
On the Sundays leading up to Spring Forward (top), peak twitter activity occurs near either 10 p.m. for states on the East Coast, or 9:30 p.m., for the rest of the country.
After Spring Forward, nearly all states exhibit peak activity later in the night. 

Looking at Texas as an individual example, before Spring Forward we see peak activity around 9:30 p.m. local time, and after Spring Forward it occurs at 11 p.m. local time.
While Texas is one of the latest peaks observed on the evening following Spring Forward, several other states are up late including Oklahoma, Georgia, and Mississippi each peaking around 10:45 p.m.

In the appendix, we show maps estimating the time of peak activity for each of the individual 9 weeks centered on Spring Forward (Figure \ref{fig:weeklymaps}). There is some week-to-week variation, most notably in the second week prior to Spring Forward, which was the night of the Academy Awards for three of the four years. By four weeks after Spring Forward, the peak activity map has relaxed to roughly the same pattern as BSF.

The magnitude of the forward shift in behavior illustrated in Figure \ref{fig:peakmap} is considered a proxy for the loss of sleep opportunity on the Sunday night following Spring Forward. 
We used two distinct methods to estimate this magnitude, namely the peak shift and the twinflection shift.
A comparison of the spatial estimates made using each method are shown in Figure \ref{fig:fourmaps}. 

Panel (a) illustrates the average shift in peak activity observed for 2011-2014 by computing the difference between the pair of maps in Figure \ref{fig:peakmap} (bottom minus top). 
While all states exhibit a shift forward in time on the night of Spring Forward, there is clear spatial variation. 
The peak in Twitter behavior for the east and west coasts occurred 15-30 minutes later Sunday night, while it occurred 45-90 minutes later for the central U.S. (Fig \ref{fig:fourmaps} panel a).

\begin{turnpage}
 \begin{figure*}
      \centering
      \includegraphics[width=1.32\textwidth]{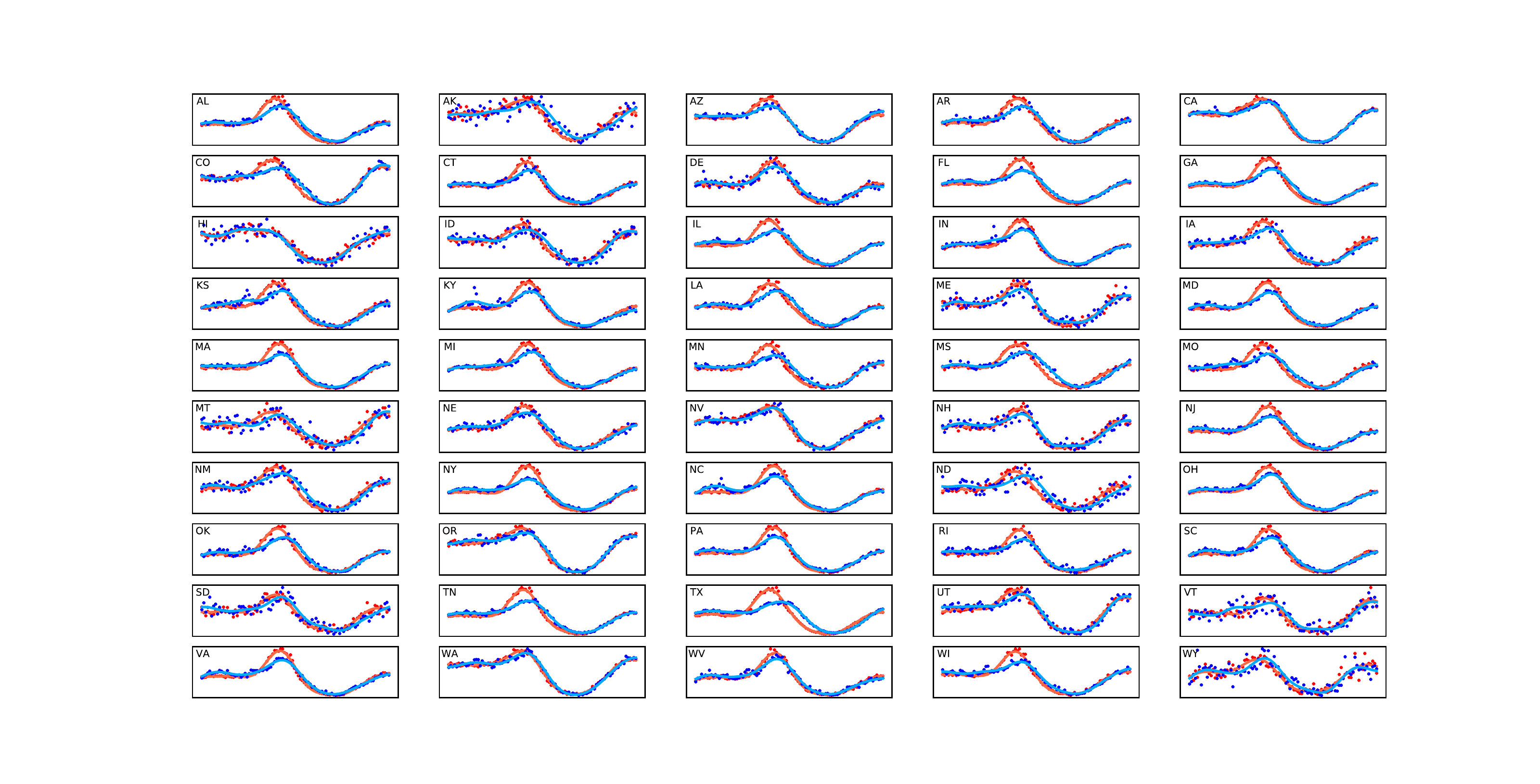}
  \caption{
  \label{fig:statecurves}
      \textbf{Normalized Twitter activity between 12 p.m. Sunday and 12 p.m. Monday prior to and following the 2014 Spring Forward event for each state.} Red indicates an aggregation of data from the specified period over four weeks before the Spring Forward Event. Blue indicates data from the single 24 hour period after Spring Forward has occurred. Dots are indicative of `raw' data, while the corresponding curves demonstrates Gaussian smoothing. Texas exhibits the largest change following Spring Forward. Curves for nearly all states have aligned by Monday morning. The BSF peaks are consistently higher than the SF peaks, largely due to televised events Before Spring Forward such as the Oscars. The Sunday of Spring Forward does not have a regularly scheduled popular television event, and as a result the SF curves have lower amplitude. %\tcm{Are there any conclusions to be drawn, not from when the peak occurs, but the difference in height? For example, HI compared to TX. CD: good question. I don't think we have the data to support much speculation on why the peak heights are different for different states, but I've added some text to the manuscript related to why the red are higher than the blue.}
      }
 \end{figure*}
 \end{turnpage}

Figure \ref{fig:fourmaps} panel (b) estimates the change using twinflection, namely the change in concavity of the behavior activity curve from down to up. 
Every state but Hawaii exhibits a shift forward in time, and with similar spatial regularity.
When measured with twinflection shift, Texas and Mississippi are seen to have the greatest temporal shift following Spring Forward. 
Texans were tweeting 135 minutes later than usual following a Spring Forward event.
Most of the east and west coast states were measured as tweeting 30 to 45 minutes later (Fig \ref{fig:fourmaps} panel b).
Both measures agreed on a positive shift for the country as a whole, and for all states exclusive of Hawaii. 
However, the two measures yielded different results for the magnitude of these shifts, with twinflection shift generally estimating a greater effect size. 

Figure \ref{fig:fourmaps} panels (c) and (d) illustrate the amount of data contributing to calculations for the behavioral curves, and the density of this data with respect to each state's population. 
Idaho, Alaska, Hawaii, Montana, Wyoming, North Dakota, South Dakota, and Vermont were the states offering the smallest amount of data, and subsequently have the highest potential for a poor behavioral curve model fit.

Though the amount of data available for California and Texas is much greater than the other states, when considering their large population size we find their twitter activity per capita to be similar to most other states. 
Based on our estimate of tweets per capita, we expect behavioral curves for most states to be more or less equally representative of their tweeting populations.

Looking at the diurnal cycle of Twitter activity for each individual state, we see remarkable consistency. 
Fig.~\ref{fig:statecurves} shows the 24 hour period spanning noon Sunday to noon Monday local time for the year 2014. 
Plots for the other 3 years exhibit similar behavior. 
Before Spring Forward (red), most states show a peak between 9:30 and 10:15 p.m., local time. 
After Spring Forward (blue), nearly all states have a peak after 10:15 p.m.
By Monday morning, nearly all curves have re-aligned. We also consistently observe higher peaks for the BSF curves which we believe to be driven by televised events such as the Oscars. The Sunday of Spring Forward does not have a regularly scheduled popular television event, and as a result the SF curves have lower amplitude.

%\tcm{a consistent shift in when the peak occurs, but why does the difference between the curve's heights differ by state?}

Both the peak and twinflection demonstrate that it is possible to observe a measurable decrease in the amount of sleep opportunity people in the United States receive on average due to Spring Forward. 
They also both demonstrate uneven geographic distribution of the effect of Spring Forward, and therefore the ability to determine geographic disparity in sleep loss.

%\tcm{could you plot a two histograms? The first histogram is the shift in the number of minutes for westcoast states and the second is a histogram of the number of minutes for east coast states } %CD: I think a histogram for all states might not be a bad idea, left of the two maps figure

\begin{figure*}[ht!]
    \centering
    \includegraphics[width=\textwidth]{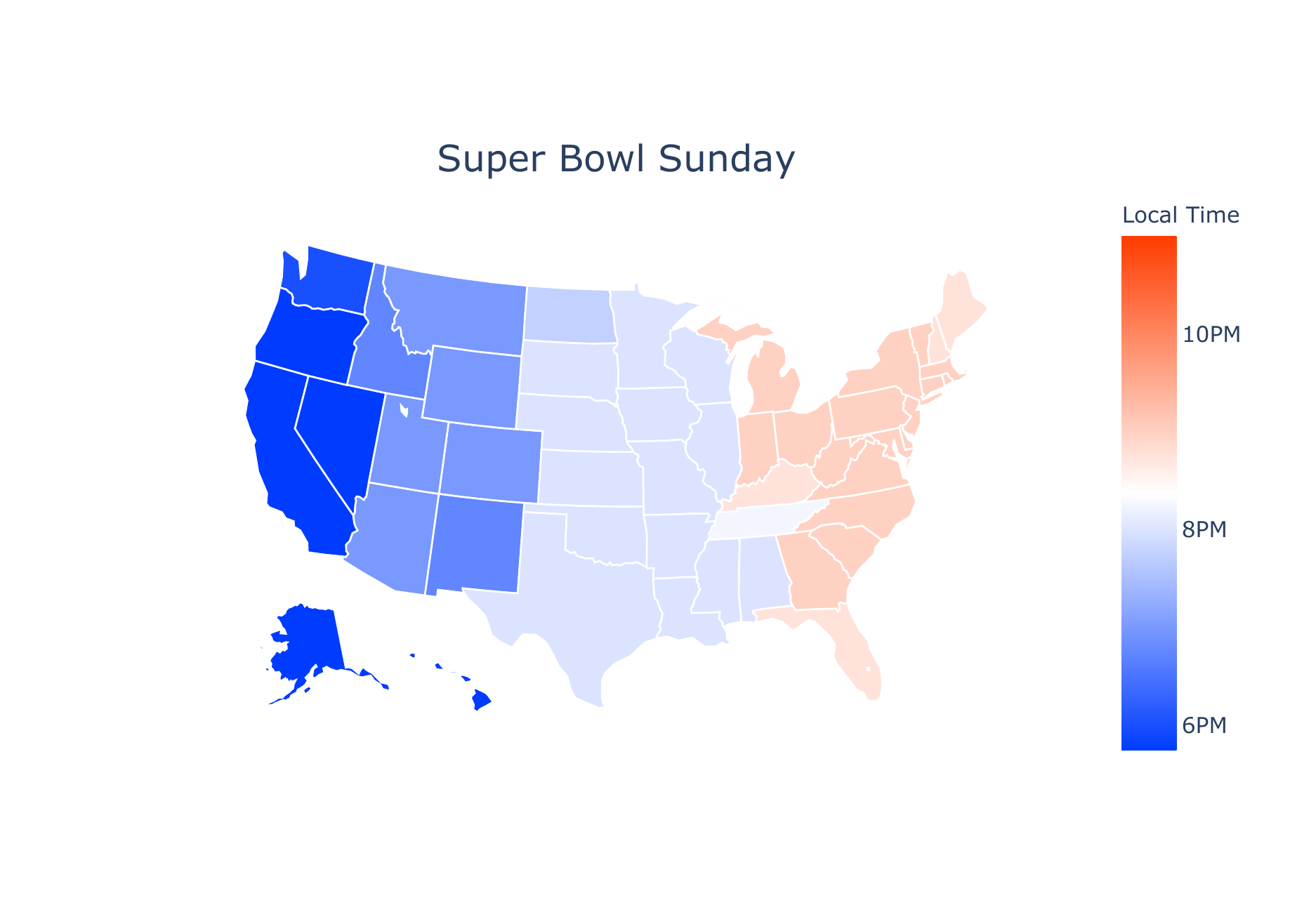}
    \caption{\textbf{Peak activity time (local) for Super Bowl Sunday, 5 weeks prior to Spring Forward, averaged over the years 2011 to 2014.} Activity exhibits a clear resemblance to the U.S. timezone map, with a peak near 9 p.m. Eastern Time just following the halftime performance. The data suggests a national collective synchronization in attention. Green Bay Packers d. Pittsburgh Steelers (2011), New York Giants d. New England Patriots (2012), Baltimore Ravens d. San Francisco 49ers (2013), and Seattle Seahawks d. Denver Broncos (2014). Performers included The Black Eyed Peas, Usher, and Slash (2011), Madonna, LMFAO, Cirque du Soleil, Nicki Minaj, M.I.A., and Cee Lo Green (2012), Beyoncé, Destiny's Child (2013), and Bruno Mars, Red Hot Chili Peppers (2014). We note that the colormap here the same as the scale used for \ref{fig:peakmap}, with blue colors included to reflect the earlier peaks seen in Mountain and Western time zones.}
    \label{fig:superbowl}
\end{figure*}

We also discovered that the Super Bowl occurred exactly 5 weeks prior to Spring Forward in each of the years studied. 
This annual event watched by over 100 million individuals in the U.S. caused peak Twitter activity to synchronize at roughly the same time nationally, around 9 p.m. Eastern, during the second half of the football game. 
The map in Figure \ref{fig:superbowl} shows the time of peak activity for each state on Super Bowl Sunday, averaged over the years 2011 to 2014. The colormap is the same as the scale used for \ref{fig:peakmap}, with the additional cooler range brought in to reflect the earlier peaks in Mountain and Western time zones.
The map bears a remarkable resemblance to the timezone map, demonstrating a synchronization of collective attention across the country.
Data from Super Bowl Sunday was not included in the Before Spring Forward data, as it does not accurately reflect the spatial distribution of typical posting behavior on a Sunday evening.

%\tcm{it is hard to see the resemblance between Fig6 and Fig3 because they use a different colormap KML: they actually don't. The color bar is the same for the later hours, but the Super Bowl peaks were much earlier than the others. So if you look at the scale NEAR 9 pm is "White" on both. I don't think it's exact bc we wouldnt be able to see half the map at all. CD: Good point Tom, I've aded a sentence in the main text and caption making explicit that the colormap is the same as that of Figure 3} 

\section{Discussion}

Technically speaking, Spring Forward occurs very early Sunday morning, and the instantaneous clock adjustment from 2 a.m. to 3 a.m. is witnessed by very few waking individuals.
In addition, we speculate that the majority of individuals do not set an alarm clock for Sunday morning.
As a result, we expect that the hour lost to Spring Forward will be felt by our bodies most meaningfully on Monday morning.
Indeed, we are likely to experience the Monday morning alarm as occurring an hour early, as Spring Forward shortens the time typically reserved for sleep opportunity Sunday night by one hour. 

Considering the correlation between screen time and lack of sleep, the Sunday evening shift, and the corresponding Monday morning re-synchronization, we observe strong evidence that sleep opportunity is lost the evening of Spring Forward.
By estimating the magnitude and spatial distribution of the shift in Twitter behavioral curves, we have approximated a lower bound on sleep loss at the state level.

Our pair of measurement methodologies have a Pearson correlation coefficient of 0.715, and a Spearman correlation coefficient of 0.645 (See Figure ~\ref{fig:correlation}). While they produced slightly different estimates of the magnitude of temporal shift in behavior, the resulting geographic profiles of sleep loss were similar. 
Both suggest that states along the coast are least affected by Spring Forward, while Texas and the states surrounding it to the North and East are the most affected. 

Peak shift suggests the temporal shift in behavior due to Spring Forward is of a similar magnitude to the actual clock shift (1 hour). 
California, the state for which we have the most data and therefore the most representative behavior profile after smoothing, was found to have a peak shift of 30 minutes. 
%\tcm{Finally the reader figures out why you plotted CA. That took a long time! CD: good noticing! Have brought this observation up into the discussion of Fig 2 thanks Tom}
Considering the clock adjustment of exactly one hour, the peak shift measurement seems likely to be directly representative of the sleep lost. Twinflection measured similar shifts for most states, but for a few estimated much larger effects. 
While California was measured as having the same 30 minute shift, Texas, the state for which we have the second most data, was estimated by twinflection to be delayed by an additional 45 minutes. 
While the relationship between magnitude of twinflection shift and magnitude of sleep loss is uncertain, this measure made spatial disparities more apparent.

Hawaii presents interesting and extreme results. 
In both cases, Hawaii is the state with the least measured sleep loss by both accounts; for twinflection shift, there is even a demonstrated gain in sleep.
Considering that Hawaii does not observe DST, these results are plausible. However, they should be considered tentative at best, given the sparsity of data available.
Caution should likewise be extended to measurements ascribed to South Dakota, North Dakota, Wyoming, Idaho, Montana, Vermont, New Hampshire, and Maine. These states have smaller populations, less population density, and lower volume of tweets. As a result, the behavioral curves associated with these states are less reliable.
%Sparse available data can lead to a greater uncertainty in establishment of a behavioral pattern, which is essential to the methodology of measurement. %\tcm{Sparse data like 1 bazillion tweets instead of 5 bazillion? The argument here feels a bit shaky? CD: Good! I've revised sentences.}

Discrepancies in available data were determined to be largely accounted for by differences in population. 
Thus, we expect results for each state (exclusive of those mentioned earlier) to be comparably reliable in their representation of sleep loss for the state as a whole.

Incremental future work in this area could include looking at the end of Daylight Savings in November, where we are ostensibly given an additional hour of sleep opportunity. 
Our findings suggest that the sleep behavior associated with other annual events including New Year's Eve and Thanksgiving ought to be visible through tweets.
More ambitiously, proxy data such as this could be verified by matching wearable measurements of sleep (e.g. Fitbit) with social media accounts.

\subsection*{Limitations}

Our study suffers from several limitations associated with our data source, we describe a few such examples here. 
The geographic location users provide in their Twitter bio is static and unlikely to be updated when traveling. 
As a result, user locations (time zone, state) inferred from this field will not always reflect their precise location. 
The GPS tagged messages included in our analysis will not suffer from this same uncertainty.
Furthermore, the tweeting population of each state is likely to have complicated biases with respect to their representation of the general population \cite{pew}. 

Our dataset likely contains automated activity. 
Indeed, an entire ecology of algorithmic tweets evolved during the period in which we collected data for this study. However, we expect the majority of this activity to be scheduled using software that updates local time automatically in response to Daylight Savings. As such, this `bot' type activity should largely serve to reduce our estimate of the time shift exhibited by humans.

As we showed for the Super Bowl, live televised events (e.g. sports, awards shows) have the potential to be a forcing mechanism to synchronize our collective attention throughout the week, and especially on Sunday evenings. 
Indeed, many individuals take to Twitter as a second screen during such events to interact with other viewers.  
In addition, streaming services such as Netflix and HBO often release new episodes of popular shows on Sunday night to align with peak consumption opportunity.
These cultural attractions exert a temporal organizing influence on our leisure behavior, and the Spring Forward disturbance translates this synchronization forward in time. 

It is worth noting that early March is a rather dull time of year for popular professional sports in the United States. 
While the National Basketball Association and National Hockey League are finishing up their regular seasons, the National Football League is in its off-season and Major League Baseball beginning pre-season exercises. 
Arguably the most engaging live-televised sporting contests taking place in early March are the NCAA College Basketball Conference Championship games, with March Madness happening weeks after Spring Forward.

In 2014, the Academy Awards were hosted by Ellen DeGeneres on Sunday March 2. Her famous selfie tweet containing many famous actors was posted that evening, a message which held the record for most retweeted status update for several years \cite{ellen}. The event happened the week before Spring Forward, and led to anomalous behavior compared with all other Sundays we looked at.

Finally, Twitter (and other social media companies) have access to much higher fidelity information regarding user activity than we have analyzed here.
We are not able to analyze consumption activity on the site, e.g. when individual messages are interacted with via views, likes, or clicks.
These forms of interaction with the Twitter ecosystem are likely to occur chronologically following the final posting of a message in the evening, and prior to the initial posting of a message in the morning. 
As a result, we expect our estimate of the sleep opportunity lost due to Spring Forward to be a lower bound.

\subsection*{Conclusion}

Privacy preserving passive measurement of daily behavior has tremendous potential to transform population-scale human activity into public health insight. 
The present study demonstrates a proof-of-concept along the path to a far more ambitious goal: construction of an `Insomniometer' capable of real-time estimation of large-scale sleep duration and quality.
Which cities in the U.S. slept well last night?
Which states are increasingly suffering from insomnia?
Answers to questions like these are not available today, but could lead to better public health surveillance in the near future. 
For example, communities exhibiting disrupted sleep in a collective pattern may be in the early stages of the outbreak of the flu or some other virus.
Current methodologies for answering these questions are not scalable, but social media, mobile devices, and wearable fitness trackers offer a new opportunity for improved monitoring of public health. 

\subsection*{Acknowledgements}
KL, TA, PSD, and CMD thank MassMutual for contributing funding in support of this research. The authors thank Adam Fox, Marc Maier, Jane Adams, David Dewhurst, Lewis Mitchell, and Henry Mitchell for helpful conversations. 

\setcounter{figure}{0}
\renewcommand{\thefigure}{A\arabic{figure}}

\setcounter{table}{0}
\renewcommand{\thetable}{A\arabic{table}}
\bibliography{bibfile}

%CD: probably fine to just leave it like this. Better to have "Appendix" not appear than to have it show up on a page by itself or be appended to end of references on the RHS column

%\appendix
%\section{Statistical details}

\begin{figure*}[!h]
 \centering
    \includegraphics[width=\textwidth]{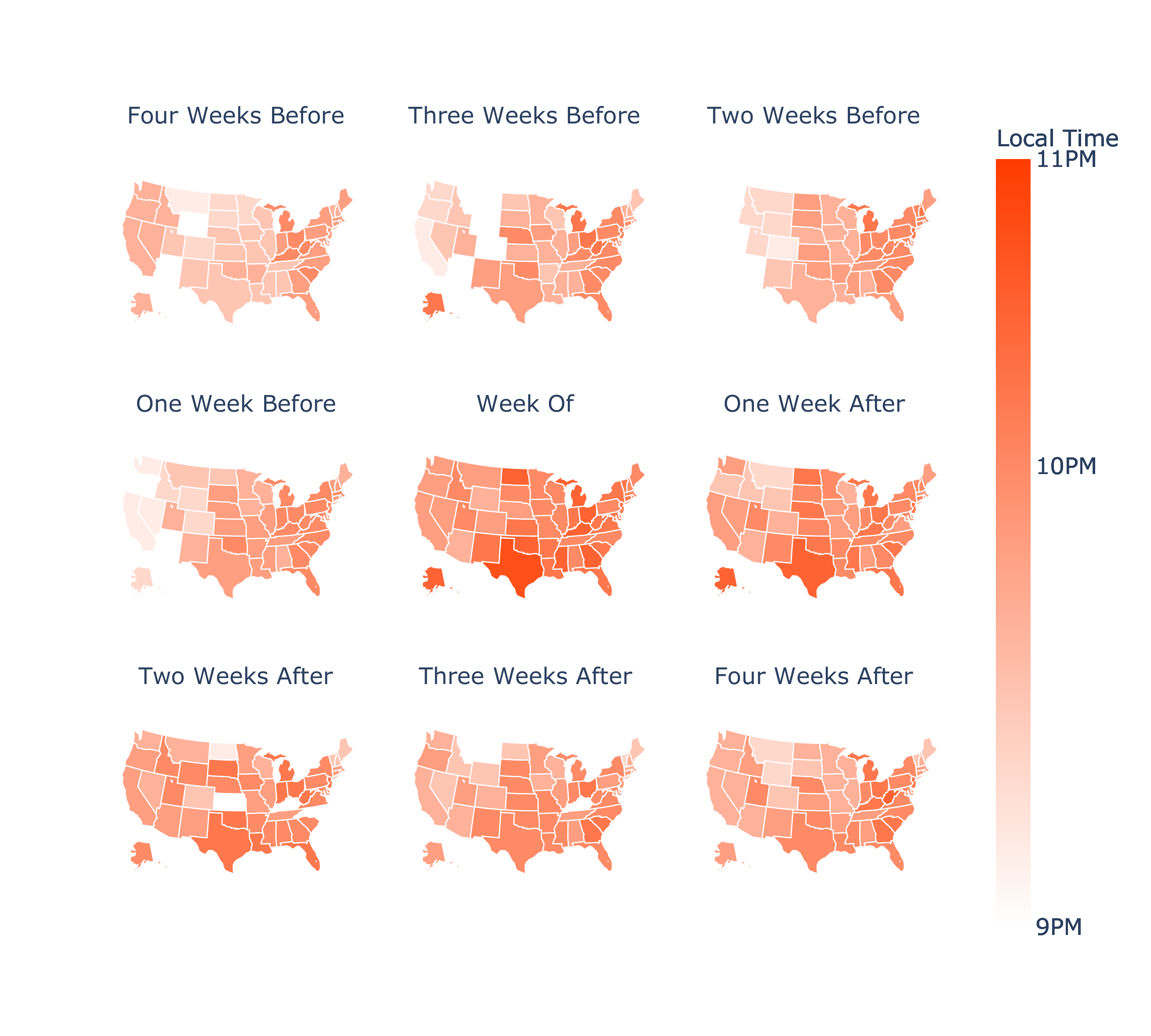}
    \caption{\textbf{Peak activity time (local) for the Sunday of the four weeks prior to, the week of, and the four weeks following Spring Forward, aggregated from 2011 to 2014.}
    We have used the same colormap as for Fig.~\ref{fig:peakmap} in the main manuscript.
    States shown in white had a peak time that was 9 pm or earlier.  From 2011 to 2013, the Academy Awards took place two weeks prior to Spring Forward, while in 2014 they took place one week prior. A clear discontinuity is visible between the ``One Week Before'' and ``Week Of'' maps.}
    \label{fig:weeklymaps}
\end{figure*}

\begin{table*}[!h]
\begin{tabular}{llllllll}
\toprule
State & count & State &  count & State & log(TPC) & State &    log(TPC) \\
\midrule
   AK &         747 &      CA &            49190 &      AK &          1.02e-03 &      DC &  2.93e-03 \\
   AL &        8072 &      TX &            45406 &      AL &          1.67e-03 &      LA &  2.35e-03 \\
   AR &        3560 &      FL &            27339 &      AR &          1.21e-03 &      DE &  2.24e-03 \\
   AZ &        6567 &      NY &            25833 &      AZ &          1.00e-03 &      MD &  1.87e-03 \\
   CA &       49190 &      OH &            21061 &      CA &          1.29e-03 &      NJ &  1.83e-03 \\
   CO &        4310 &      PA &            18790 &      CO &          8.31e-04 &      OH &  1.82e-03 \\
   CT &        5263 &      MI &            17259 &      CT &          1.47e-03 &      TX &  1.81e-03 \\
   DC &        1854 &      IL &            16620 &      DC &          2.93e-03 &      RI &  1.76e-03 \\
   DE &        2051 &      NJ &            16216 &      DE &          2.24e-03 &      MI &  1.75e-03 \\
   FL &       27339 &      GA &            15952 &      FL &          1.42e-03 &      NV &  1.74e-03 \\
   GA &       15952 &      NC &            13600 &      GA &          1.61e-03 &      AL &  1.67e-03 \\
   HI &        1309 &      VA &            11761 &      HI &          9.40e-04 &      SC &  1.65e-03 \\
   IA &        4233 &      MD &            11030 &      IA &          1.38e-03 &      GA &  1.61e-03 \\
   ID &         934 &      LA &            10822 &      ID &          5.85e-04 &      MA &  1.50e-03 \\
   IL &       16620 &      MA &             9995 &      IL &          1.29e-03 &      PA &  1.47e-03 \\
   IN &        8138 &      TN &             8173 &      IN &          1.24e-03 &      CT &  1.47e-03 \\
   KS &        4063 &      IN &             8138 &      KS &          1.41e-03 &      KY &  1.45e-03 \\
   KY &        6373 &      AL &             8072 &      KY &          1.45e-03 &      WV &  1.45e-03 \\
   LA &       10822 &      SC &             7817 &      LA &          2.35e-03 &      OK &  1.44e-03 \\
   MA &        9995 &      WA &             7469 &      MA &          1.50e-03 &      VA &  1.44e-03 \\
   MD &       11030 &      AZ &             6567 &      MD &          1.87e-03 &      FL &  1.42e-03 \\
   ME &         965 &      KY &             6373 &      ME &          7.26e-04 &      KS &  1.41e-03 \\
   MI &       17259 &      MO &             6099 &      MI &          1.75e-03 &      MS &  1.40e-03 \\
   MN &        5258 &      WI &             5705 &      MN &          9.77e-04 &      NC &  1.39e-03 \\
   MO &        6099 &      OK &             5495 &      MO &          1.01e-03 &      IA &  1.38e-03 \\
   MS &        4182 &      CT &             5263 &      MS &          1.40e-03 &      NY &  1.32e-03 \\
   MT &         369 &      MN &             5258 &      MT &          3.67e-04 &      CA &  1.29e-03 \\
   NC &       13600 &      NV &             4792 &      NC &          1.39e-03 &      IL &  1.29e-03 \\
   ND &         780 &      CO &             4310 &      ND &          1.11e-03 &      TN &  1.25e-03 \\
   NE &        2262 &      IA &             4233 &      NE &          1.22e-03 &      IN &  1.24e-03 \\
   NH &        1128 &      MS &             4182 &      NH &          8.54e-04 &      NE &  1.22e-03 \\
   NJ &       16216 &      KS &             4063 &      NJ &          1.83e-03 &      AR &  1.21e-03 \\
   NM &        1846 &      OR &             3871 &      NM &          8.85e-04 &      ND &  1.11e-03 \\
   NV &        4792 &      AR &             3560 &      NV &          1.74e-03 &      WA &  1.08e-03 \\
   NY &       25833 &      WV &             2683 &      NY &          1.32e-03 &      AK &  1.02e-03 \\
   OH &       21061 &      UT &             2495 &      OH &          1.82e-03 &      MO &  1.01e-03 \\
   OK &        5495 &      NE &             2262 &      OK &          1.44e-03 &      AZ &  1.00e-03 \\
   OR &        3871 &      DE &             2051 &      OR &          9.93e-04 &      WI &  9.96e-04 \\
   PA &       18790 &      DC &             1854 &      PA &          1.47e-03 &      OR &  9.93e-04 \\
   RI &        1845 &      NM &             1846 &      RI &          1.76e-03 &      MN &  9.77e-04 \\
   SC &        7817 &      RI &             1845 &      SC &          1.65e-03 &      HI &  9.40e-04 \\
   SD &         540 &      HI &             1309 &      SD &          6.48e-04 &      NM &  8.85e-04 \\
   TN &        8173 &      NH &             1128 &      TN &          1.25e-03 &      UT &  8.74e-04 \\
   TX &       45406 &      ME &              965 &      TX &          1.81e-03 &      NH &  8.54e-04 \\
   UT &        2495 &      ID &              934 &      UT &          8.74e-04 &      CO &  8.31e-04 \\
   VA &       11761 &      ND &              780 &      VA &          1.44e-03 &      VT &  7.59e-04 \\
   VT &         475 &      AK &              747 &      VT &          7.59e-04 &      ME &  7.26e-04 \\
   WA &        7469 &      SD &              540 &      WA &          1.08e-03 &      SD &  6.48e-04 \\
   WI &        5705 &      VT &              475 &      WI &          9.96e-04 &      ID &  5.85e-04 \\
   WV &        2683 &      MT &              369 &      WV &          1.45e-03 &      WY &  4.25e-04 \\
   WY &         245 &      WY &              245 &      WY &          4.25e-04 &      MT &  3.67e-04 \\
\bottomrule
\end{tabular}
  \caption{\textbf{Tweet Counts.}
   Tweet count and tweets per capita ($\log_{10}$) sorted alphabetically and in order of volume for the four ASF Sundays observed in 2011-2014.}
    \label{tab:datatable}
\end{table*}

\begin{table*}
\begin{tabular}{llllllll}
\toprule
State  &  BSF & State  & BSF & State & SF & State &  SF \\
\midrule
     AK &       09:45 &      PA &    10:15 &    AK &      10:45 &      TX &    11:00 \\
     AL &       09:30 &      FL &    10:15 &    AL &      10:15 &      AK &    10:45 \\
     AR &       09:45 &      NY &    10:15 &    AR &      10:30 &      GA &    10:45 \\
     AZ &       09:30 &      KY &    10:15 &    AZ &      09:45 &      OK &    10:45 \\
     CA &       09:30 &      OH &    10:15 &    CA &      10:00 &      OH &    10:45 \\
     CO &       09:00 &      IN &    10:15 &    CO &      10:00 &      ND &    10:45 \\
     CT &       10:00 &      MI &    10:15 &    CT &      10:30 &      MI &    10:45 \\
     DC &       10:00 &      GA &    10:15 &    DC &      10:00 &      KY &    10:45 \\
     DE &       10:00 &      SC &    10:15 &    DE &      10:15 &      MS &    10:45 \\
     FL &       10:15 &      NJ &    10:15 &    FL &      10:30 &      FL &    10:30 \\
     GA &       10:15 &      VA &    10:15 &    GA &      10:45 &      LA &    10:30 \\
     HI &       06:00 &      WV &    10:15 &    HI &      06:00 &      NM &    10:30 \\
     IA &       09:30 &      DE &    10:00 &    IA &      10:15 &      NY &    10:30 \\
     ID &       09:30 &      DC &    10:00 &    ID &      10:15 &      RI &    10:30 \\
     IL &       09:30 &      CT &    10:00 &    IL &      10:15 &      SC &    10:30 \\
     IN &       10:15 &      RI &    10:00 &    IN &      10:30 &      CT &    10:30 \\
     KS &       09:45 &      NC &    10:00 &    KS &      10:30 &      MD &    10:30 \\
     KY &       10:15 &      NH &    10:00 &    KY &      10:45 &      AR &    10:30 \\
     LA &       09:30 &      MA &    10:00 &    LA &      10:30 &      KS &    10:30 \\
     MA &       10:00 &      MD &    10:00 &    MA &      10:15 &      IN &    10:30 \\
     MD &       10:00 &      ME &    10:00 &    MD &      10:30 &      VA &    10:30 \\
     ME &       10:00 &      NM &    09:45 &    ME &      10:15 &      VT &    10:30 \\
     MI &       10:15 &      AK &    09:45 &    MI &      10:45 &      WV &    10:30 \\
     MN &       09:30 &      OK &    09:45 &    MN &      10:15 &      NJ &    10:30 \\
     MO &       09:30 &      TN &    09:45 &    MO &      10:15 &      UT &    10:15 \\
     MS &       09:45 &      VT &    09:45 &    MS &      10:45 &      DE &    10:15 \\
     MT &       09:00 &      NE &    09:45 &    MT &      10:00 &      TN &    10:15 \\
     NC &       10:00 &      MS &    09:45 &    NC &      10:15 &      SD &    10:15 \\
     ND &       09:30 &      AR &    09:45 &    ND &      10:45 &      PA &    10:15 \\
     NE &       09:45 &      KS &    09:45 &    NE &      10:00 &      WI &    10:15 \\
     NH &       10:00 &      ND &    09:30 &    NH &      10:15 &      NH &    10:15 \\
     NJ &       10:15 &      SD &    09:30 &    NJ &      10:30 &      MN &    10:15 \\
     NM &       09:45 &      WI &    09:30 &    NM &      10:30 &      ME &    10:15 \\
     NV &       09:30 &      WA &    09:30 &    NV &      10:00 &      IA &    10:15 \\
     NY &       10:15 &      AZ &    09:30 &    NY &      10:30 &      NC &    10:15 \\
     OH &       10:15 &      CA &    09:30 &    OH &      10:45 &      ID &    10:15 \\
     OK &       09:45 &      UT &    09:30 &    OK &      10:45 &      IL &    10:15 \\
     OR &       09:30 &      TX &    09:30 &    OR &      10:00 &      AL &    10:15 \\
     PA &       10:15 &      IA &    09:30 &    PA &      10:15 &      MO &    10:15 \\
     RI &       10:00 &      ID &    09:30 &    RI &      10:30 &      MA &    10:15 \\
     SC &       10:15 &      OR &    09:30 &    SC &      10:30 &      WA &    10:00 \\
     SD &       09:30 &      IL &    09:30 &    SD &      10:15 &      DC &    10:00 \\
     TN &       09:45 &      LA &    09:30 &    TN &      10:15 &      NE &    10:00 \\
     TX &       09:30 &      NV &    09:30 &    TX &      11:00 &      CO &    10:00 \\
     UT &       09:30 &      MN &    09:30 &    UT &      10:15 &      MT &    10:00 \\
     VA &       10:15 &      MO &    09:30 &    VA &      10:30 &      OR &    10:00 \\
     VT &       09:45 &      AL &    09:30 &    VT &      10:30 &      CA &    10:00 \\
     WA &       09:30 &      MT &    09:00 &    WA &      10:00 &      NV &    10:00 \\
     WI &       09:30 &      CO &    09:00 &    WI &      10:15 &      AZ &    09:45 \\
     WV &       10:15 &      WY &    09:00 &    WV &      10:30 &      WY &    09:45 \\
     WY &       09:00 &      HI &    06:00 &    WY &      09:45 &      HI &    06:00 \\
\bottomrule
\end{tabular}
\caption{\textbf{Time of Peak Twitter Activity by State.}
Time of peak Twitter activity Before Spring Forward (BSF) and the week of Spring Forward (SF) for each state, listed alphabetically and by time of peak.}
\label{tab:peaktable}
\end{table*}

\begin{table*}
 \begin{tabular}{llllllll}
\toprule
State &  Peak  & State &  Peak  & State &  Twin & State &  Twin   \\
\midrule
       AK &          60 &      TX &           90 &    AK &          60 &      TX &          135 \\
     AL &          45 &      ND &           75 &    AL &          60 &      MS &          105 \\
     AR &          45 &      AK &           60 &    AR &          60 &      LA &           90 \\
     AZ &          15 &      LA &           60 &    AZ &          30 &      ID &           75 \\
     CA &          30 &      OK &           60 &    CA &          30 &      TN &           75 \\
     CO &          60 &      MT &           60 &    CO &          75 &      CO &           75 \\
     CT &          30 &      MS &           60 &    CT &          30 &      ND &           75 \\
     DC &           0 &      CO &           60 &    DC &          45 &      MN &           75 \\
     DE &          15 &      WY &           45 &    DE &          15 &      IL &           75 \\
     FL &          15 &      MO &           45 &    FL &          45 &      WI &           60 \\
     GA &          30 &      MN &           45 &    GA &          45 &      OK &           60 \\
     HI &           0 &      UT &           45 &    HI &         -30 &      NM &           60 \\
     IA &          45 &      AR &           45 &    IA &          45 &      AL &           60 \\
     ID &          45 &      VT &           45 &    ID &          75 &      AK &           60 \\
     IL &          45 &      SD &           45 &    IL &          75 &      AR &           60 \\
     IN &          15 &      KS &           45 &    IN &          30 &      IA &           45 \\
     KS &          45 &      NM &           45 &    KS &          45 &      MO &           45 \\
     KY &          30 &      IL &           45 &    KY &          45 &      VT &           45 \\
     LA &          60 &      ID &           45 &    LA &          90 &      UT &           45 \\
     MA &          15 &      IA &           45 &    MA &          45 &      SC &           45 \\
     MD &          30 &      WI &           45 &    MD &          45 &      OH &           45 \\
     ME &          15 &      AL &           45 &    ME &          15 &      NJ &           45 \\
     MI &          30 &      TN &           30 &    MI &          45 &      NE &           45 \\
     MN &          45 &      CA &           30 &    MN &          75 &      FL &           45 \\
     MO &          45 &      NV &           30 &    MO &          45 &      DC &           45 \\
     MS &          60 &      OH &           30 &    MS &         105 &      GA &           45 \\
     MT &          60 &      MI &           30 &    MT &          30 &      KS &           45 \\
     NC &          15 &      OR &           30 &    NC &          30 &      MI &           45 \\
     ND &          75 &      MD &           30 &    ND &          75 &      KY &           45 \\
     NE &          15 &      CT &           30 &    NE &          45 &      MD &           45 \\
     NH &          15 &      KY &           30 &    NH &          30 &      MA &           45 \\
     NJ &          15 &      WA &           30 &    NJ &          45 &      MT &           30 \\
     NM &          45 &      GA &           30 &    NM &          60 &      WA &           30 \\
     NV &          30 &      RI &           30 &    NV &          30 &      VA &           30 \\
     NY &          15 &      VA &           15 &    NY &          30 &      AZ &           30 \\
     OH &          30 &      SC &           15 &    OH &          45 &      CA &           30 \\
     OK &          60 &      WV &           15 &    OK &          60 &      SD &           30 \\
     OR &          30 &      NE &           15 &    OR &          30 &      RI &           30 \\
     PA &           0 &      AZ &           15 &    PA &          30 &      PA &           30 \\
     RI &          30 &      NY &           15 &    RI &          30 &      OR &           30 \\
     SC &          15 &      NJ &           15 &    SC &          45 &      CT &           30 \\
     SD &          45 &      NH &           15 &    SD &          30 &      NY &           30 \\
     TN &          30 &      DE &           15 &    TN &          75 &      NV &           30 \\
     TX &          90 &      NC &           15 &    TX &         135 &      IN &           30 \\
     UT &          45 &      ME &           15 &    UT &          45 &      NH &           30 \\
     VA &          15 &      MA &           15 &    VA &          30 &      NC &           30 \\
     VT &          45 &      IN &           15 &    VT &          45 &      ME &           15 \\
     WA &          30 &      FL &           15 &    WA &          30 &      DE &           15 \\
     WI &          45 &      PA &            0 &    WI &          60 &      WV &           15 \\
     WV &          15 &      HI &            0 &    WV &          15 &      WY &           15 \\
     WY &          45 &      DC &            0 &    WY &          15 &      HI &          -30 \\
\bottomrule
\end{tabular}
\caption{ \textbf{Spring Forward Time Shift (minutes) by State.}
The temporal shift in (1) peak activity and (2) twinflection sorted alphabetically and by magnitude. Times reported are differences between columns in the preceding table, and reported in minutes.}
\label{tab:shifttable}
\end{table*}

\begin{figure*} 
 \centering
    \includegraphics[width=.75\textwidth]{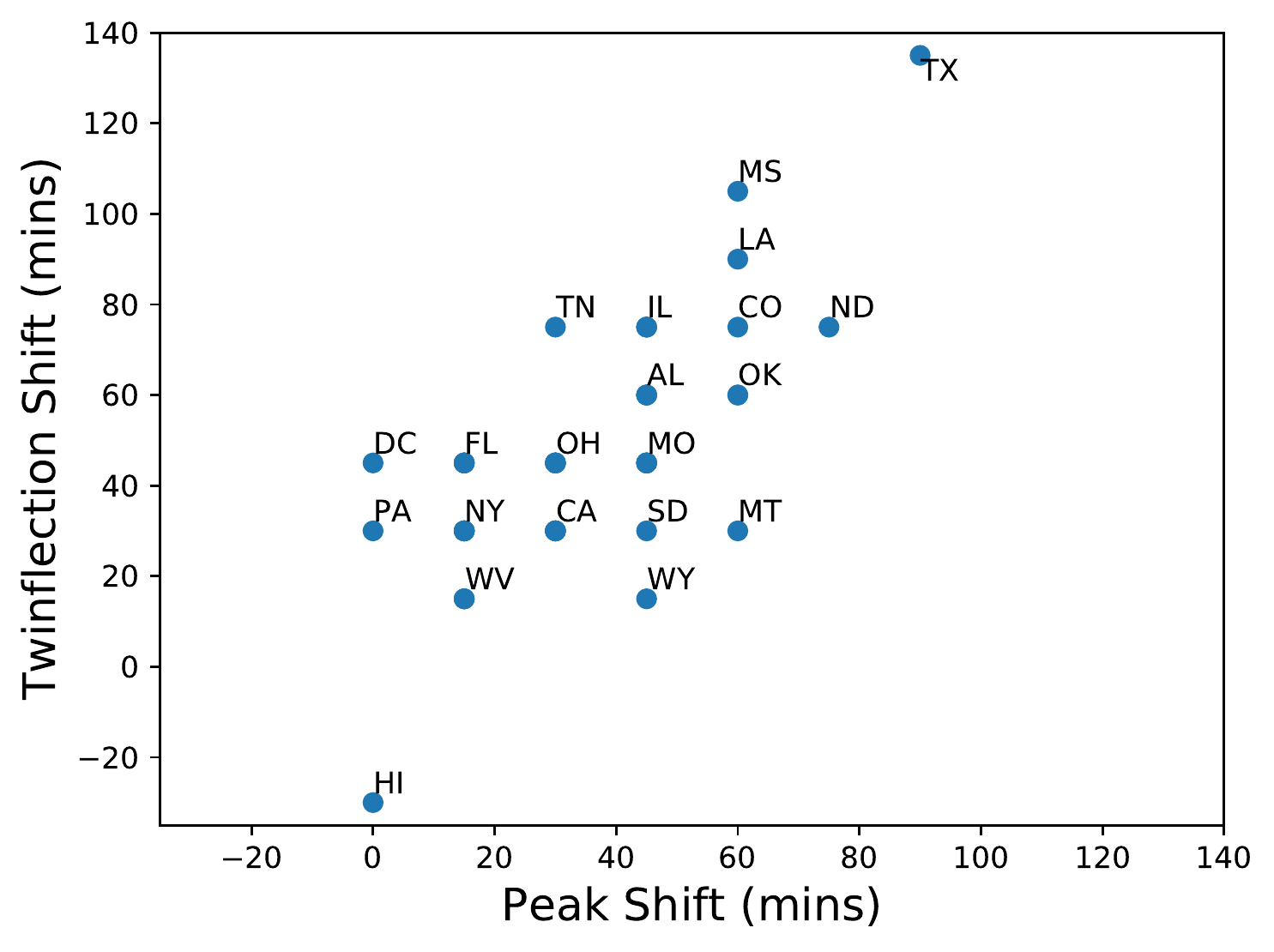}
    \caption{\textbf{Correlation of Peak and Twinflection shift estimates.} Blue discs represent one or more states having that combination of ordered pair estimates (peak shift, twinflection shift). State abbreviations label each comparison. Given that there is overlap, we label each concurrent point with the state contributing the greatest number of tweets. Table ~\ref{tab:shifttable} reports all states and shifts using each measure. The Pearson correlation of the two measures plotted here is 0.715, while the Spearman rank correlation is 0.645. }

    \label{fig:correlation}
\end{figure*}

\end{document}